\documentclass[sigconf]{acmart}

\usepackage{array}
\AtBeginDocument{%
  \providecommand\BibTeX{{%
    \normalfont B\kern-0.5em{\scshape i\kern-0.25em b}\kern-0.8em\TeX}}}

\copyrightyear{2021}
\acmYear{2021}
\setcopyright{acmcopyright}\acmConference[SIGMOD '21]{Proceedings of the 2021 International Conference on Management of Data}{June 20--25, 2021}{Virtual Event, China}
\acmBooktitle{Proceedings of the 2021 International Conference on Management of Data (SIGMOD '21), June 20--25, 2021, Virtual Event, China}
\acmPrice{15.00}
\acmDOI{10.1145/3448016.3457552}
\acmISBN{978-1-4503-8343-1/21/06}

\begin{document}
\fancyhead{}
\title{Real-time Data Infrastructure at Uber}

\author{Yupeng Fu}
\email{yupeng@uber.com}
\affiliation{%
  \institution{Uber, Inc}
}

\author{Chinmay Soman}
\email{chinmay.cerebro@gmail.com}
\affiliation{%
  \institution{Uber, Inc}
}

\renewcommand{\shortauthors}{Fu, et al.}

\begin{abstract}
Uber’s business is highly real-time in nature. PBs of data is continuously being collected from the end users such as Uber drivers, riders, restaurants, eaters and so on everyday. There is a lot of valuable information to be processed and many decisions must be made in seconds for a variety of use cases such as customer incentives, fraud detection, machine learning model prediction. In addition, there is an increasing need to expose this ability to different user categories, including engineers, data scientists, executives and operations personnel which adds to the complexity. 

In this paper, we present the overall architecture of the real-time data infrastructure and identify three scaling challenges that we need to continuously address for each component in the architecture. At Uber, we heavily rely on open source technologies for the key areas of the infrastructure. On top of those open-source software, we add significant improvements and customizations to make the open-source solutions fit in Uber’s environment and bridge the gaps to meet Uber’s unique scale and requirements.

We then highlight several important use cases and show their real-time solutions and tradeoffs. Finally, we reflect on the lessons we learned as we built, operated and scaled these systems.

\end{abstract}



\maketitle

\section{Introduction} \label{sec:intro}
A lot of real-time data is generated within Uber’s data centers. This data originates in different sources such as end user applications (driver/rider/eater) or the backend microservices. Some of this data consists of application or system logs continuously emitted as part of day to day operation. A lot of services also emit special events for tracking things such as trip updates, driver status change, order cancellation and so on. Some of it is also derived from the OnLine Transactional Processing (OLTP) database changelog used internally by such microservices. As of October 2020, trillions of messages and petabytes of such data were generated per day across all regions. 

Real-time data processing plays a critical role in Uber’s technology stack and it empowers a wide range of use cases. At high level, real-time data processing needs within Uber consists of three broad areas: 1) Messaging platform that allows communication between asynchronous producers and subscribers 2) Stream processing that allows applying computational logic on top of such streams of messages and 3) OnLine Analytical Processing (OLAP) that enables analytical queries over all this data in near real time. Each area has to deal with three fundamental scaling challenges within Uber:

\begin{itemize}
\item Scaling data: The total incoming real time data volume has been growing exponentially at a rapid rate of year over year produced by several thousands of micro services. In addition, Uber deploys its infrastructure in several geographical regions for high availability, and it has a multiplication factor in terms of handling aggregate data. Each real time processing system has to handle this data volume increase while maintaining SLA around data freshness, end-to-end latency and availability.

\item Scaling use cases: As Uber’s business grows, new use cases emerge from various business verticals and groups. Different parts of the organization have varying requirements for the real time data systems, which are often competing in nature. For instance, dynamic pricing\cite{chen2016dynamic} for a given Uber product (such as rides or eats) is a highly complex real-time workflow involving multi-stage stream processing pipelines that run various machine learning algorithms along with a fast key-value store. This system is designed for favoring freshness and availability over data consistency, and it’s implemented entirely by engineers. On the other hand, monitoring real-time business metrics around orders and sales requires a SQL like interface used by data scientists with more emphasis given to data completeness. 

\item Scaling users: The diverse users interacting with the real time data system fall on a big spectrum of technical skills from operations personnel who have no engineering background to advanced users capable of orchestrating complex real time computational data pipelines. As the personnel of Uber grows, the platform teams also face increasing challenges on the user imposed complexities, such as safe client-side version upgrade for a large number of applications and managing an increasing number of user requests.

\end{itemize}

In short, the biggest challenge is to build a unified platform with standard abstractions that can work for all such varied use cases and users at scale instead of creating custom solutions. A key decision made by us to overcome such challenges is to adopt the open source solutions in building this unified platform. Open source software adoption has many advantages such as the development velocity, cost effectiveness as well as the power of the crowd. Given the scale and rapid development cycles in Uber, we had to pick technologies that were mature enough to be able to scale with Uber’s data as well as extensible enough for us to integrate it in our unified real-time data stack. 

\begin{figure}[h]
  \centering
  \includegraphics[width=\linewidth]{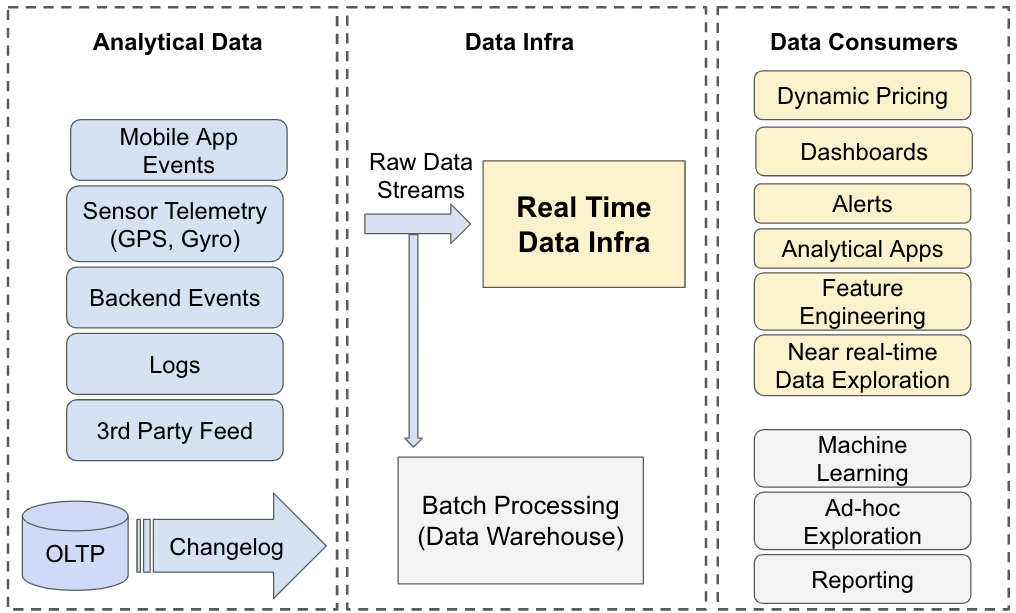}
  \caption{The high-level data flow at Uber infrastructure}
  \label{fig:data-flow}
\end{figure}

Figure \ref{fig:data-flow} depicts the high-level flow of data inside Uber's infrastructure. Various kinds of analytical data are continuously collected from Uber’s data centers across multiple regions. These streams of raw data form the source of truth for all analytics at Uber. Most of these streams are incrementally archived in batch processing systems and ingested in the data warehouse. This is then made available for machine learning and other data science use cases. The Real Time Data Infra component continuously processes such data streams for powering a variety of mission critical use cases such as dynamic pricing (Surge), intelligent alerting, operational dashboards and so on. This paper focuses on the real time data eco-system. 

The paper is organized as follows. In Section \ref{sec:requirements}, we list the requirements derived from the use cases which are used to guide the design decisions for various real-time use cases. In Section \ref{sec:abstractions}, we provide an overview of the high-level abstractions of the real-time data infrastructure at Uber. In Section \ref{sec:overview}, we present the open source technologies that we adopted for each component in the architecture. More importantly, we describe the enhancements and improvements on the open source solutions to overcome the scaling challenges faced by Uber. In Section \ref{sec:use-case}, we analyze several real-time use cases at Uber and show how their solutions are shaped differently due to unique design requirements. We then discuss a few other important aspects of the real-time data infrastructure in Section \ref{sec:all-active} and Section \ref{sec:backfill}. Then in Section \ref{sec:related} we discuss related work and in Section \ref{sec:lessons}, we reflect on lessons we learned about building and operating real-time systems at Uber. Finally, we conclude in Section \ref{sec:conclusion} and show the future work in Section \ref{sec:future}.

\section{Requirements} \label{sec:requirements}

Each category of use case mentioned in Section \ref{sec:intro} has its own special requirements pertaining to real-time data infrastructure which are often competing with those of other use cases. The different requirements generally include the following aspects:

\begin{itemize}
\item Consistency: Mission-critical applications such as financial dashboards require data to be consistent across all regions. This includes zero data loss in the inter-region and intra-region dispersal and processing mechanisms, de-duplication as well as ability to certify data quality. 

\item Availability: The real time data infrastructure stack must be highly available with 99.99 percentile guarantee. Loss of availability has a direct impact on Uber’s business and may result in significant financial losses. For instance, dynamic pricing leverages the real-time data infrastructure component for calculating demand and supply ratios per geo-fence, which in turn is used to influence the price of a trip or UberEats delivery.

\item Data Freshness: Most of the use cases require seconds level freshness. In other words a given event or log record must be available for processing or querying, seconds after it has been produced. This is a critical requirement to ensure ability to respond to certain events such as security incidents, demand-supply skews, business metric alerts and so on.

\item Query latency: Some use cases need the ability to execute queries on the raw data stream and require the p99th query latency to be under 1 second. For instance, site facing or external analytical tools such as UberEats Restaurant manager\cite{rm} will execute several analytical queries for each page load. Each such query must be very fast to provide a good experience for the restaurant owner.

\item  Scalability: The raw data streams constitute petabytes of data volume collected per day across all regions. This data is constantly growing based on organic growth of our user base, new lines of business deployed by Uber as well as new real time analytics use cases that arise over time. The ability to scale with this ever growing data set in a seamless manner, without requiring users to re-architect the processing pipelines is a fundamental requirement of the real-time data infrastructure stack.

\item  Cost: Uber is a low margin business. We need to ensure the cost of data processing and serving is low and ensure high operational efficiency. This influences a variety of design decisions such as amount of data kept in memory, tiered storage, pre-materialization vs runtime computation and so on.

\item Flexibility: We need to provide programmatic as well as declarative (SQL like) interface for expressing computational logic to accommodate the diverse user groups. In addition, some use cases need a push-based model which is semi-stateful and continuously emits generated results whereas others might need a stateful pull-based model where the user can execute queries on the raw data stream. For instance, users can create intelligent alerts in case of business rule violation using push-based stream processing pipelines. Whereas, dashboarding and triaging will require a pull-based SQL interface for the same datasets.
\end{itemize}

It’s easy to observe that guaranteeing all these requirements for the same use case is not possible. For instance in the dynamic pricing use case, we cannot guarantee both data consistency and freshness (availability) at Uber’s scale based on CAP theorem\cite{gilbert2002brewer}. To minimize business impact, we must therefore prioritize freshness ahead of consistency. Subsequently, each technology chosen for building this use case must be finely tuned for favouring freshness. Such tradeoffs are discussed in detail in Section \ref{sec:use-case} by analyzing several real-time use cases at Uber.

\section{Abstractions} \label{sec:abstractions}

The diagram in Figure \ref{fig:abstractions} illustrates the logical building blocks that constitute a real-time analytics stack. The different components (from bottom up) are as follows:

\begin{figure}[h]
  \centering
  \includegraphics[width=\linewidth]{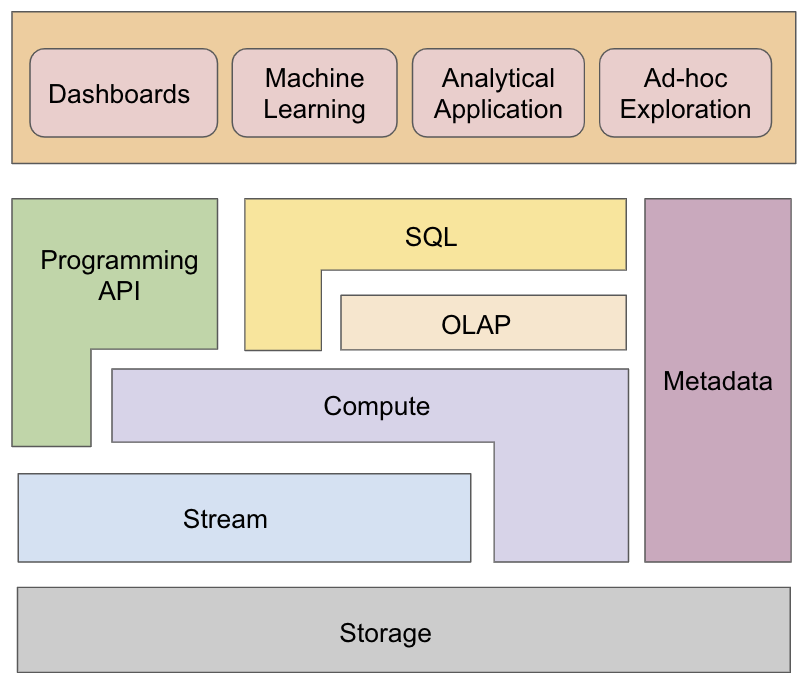}
  \caption{An abstraction of the real-time data infrastructure and the overview of the components}
  \label{fig:abstractions}
\end{figure}

{\bfseries Storage.} This provides a generic object or blob storage interface for all the layers above it with a read after write consistency guarantee. This is primarily used for long term storage of data and should be optimized for high write rate. Reads are less frequent and used for cases such as bootstrapping data in an OLAP table or a stream, data backfills and so on.

{\bfseries Stream.} This provides a publish-subscribe interface to the higher layers. Users of this layer can produce events to a particular stream or topic. Any other user subscribing to this stream can consume the data one event at a time. This system should be optimized for low latency for reads and writes. The minimum requirements from this layer include ability to partition the data and at least once semantics between producer and subscriber.

{\bfseries Compute.} This provides the ability to perform arbitrary computation on the underlying stream and storage layers. When computing over a stream, processing happens for each individual event whereas computation done directly over storage can be done in batches. It’s important to note that we can choose the same or different technologies for stream processing vs storage or batch processing. Choosing the same technology will result in simpler abstraction for the higher layers but results in higher complexity to implement - thus adding to significant operational overhead. Whereas, choosing two different technologies makes the individual components feasible but delegates the task of federation to the higher layers. The minimum requirements from this layer include at least once semantics between the data source and sink.

{\bfseries OLAP.} This layer provides a limited SQL capability over data coming from stream or storage. The system should be optimized for serving analytical queries including filtering, aggregations with group by, order by in a high throughput, low latency manner. The minimum requirements from this layer for the vast majority of use cases include at least once semantics while ingesting data from the different sources. Exactly once data ingestion based on a primary key is a must have for a small set of critical use cases. 

{\bfseries SQL.} This refers to a full SQL query layer on top of OLAP as well as compute. When used with the compute layer, the SQL statement is compiled into a compute function which can be applied to the underlying stream or storage. When used with the OLAP layer, it will do additional processing on top of the limited SQL provided, to fill in the gaps. For instance, most real-time OLAP databases have limited or no join support, and this can be done at this SQL layer. It’s interesting to note that joins can also be done in lower layers (pre-materialize at the compute layer) and served by the OLAP layer without need for additional processing - albeit at a higher cost. The minimum requirements from this layer include SQL semantics which are closer to ANSI SQL with extensions applicable for stream processing (for instance - window functions).

{\bfseries API.} This provides a programmatic way to access the stream or specify a compute function for the higher layer applications. This is to be used by advanced users for whom the SQL interface is not sufficient. It’s important to note that the choice of technologies in the layers below will have a direct impact on the simplicity of this API. 

{\bfseries Metadata.} This provides a simple interface to manage all kinds of metadata required for all the aforementioned layers. For instance, the schema that describes structured data managed by storage or stream will be stored here. Minimum requirements include ability to version the metadata and have checks for ensuring backward compatibility across versions.

\section{System Overview} \label{sec:overview}

\begin{figure}[h]
  \centering
  \includegraphics[width=\linewidth]{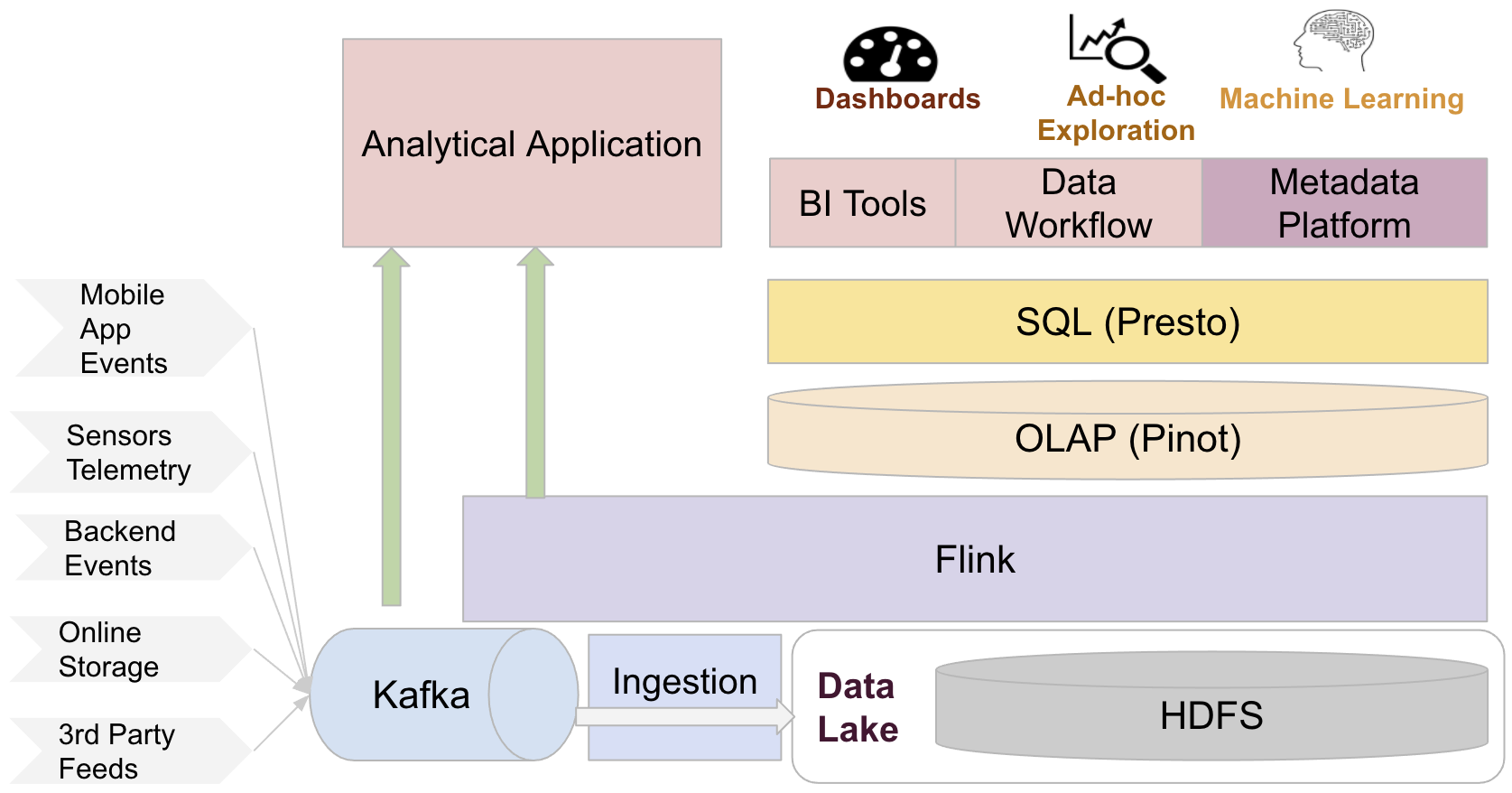}
  \caption{Overview of the real-time data infrastructure at Uber}
  \label{fig:overview}
\end{figure}

Each following subsection introduces the open source systems we have adopted for the corresponding logical building block as shown in Figure \ref{fig:overview}. We then subsequently describe Uber’s unique contributions in each domain and explain how it bridges the gaps to meet Uber’s unique scale and requirements.

\subsection{Apache Kafka for streaming storage} \label{sec:kafka}

Kafka\cite{kreps2011kafka} is a popular open-source distributed event streaming system that is widely used in the industry. In 2015 Kafka was already a popular solution known for good performance when we adopted it. A more recent performance study can be found in this benchmark report from Confluent\cite{confluent-test}, which compared Kafka, Pulsar\cite{pulsar} and RabbitMQ\cite{rabbitmq} on system throughput and latency, the primary performance metrics for event streaming systems in production. Besides the performance, there were several other important factors to consider for adoption, such as operational simplicity, open-source ecosystem maturity, size of open-source community, adoption rate across the industry. Looking at those together, Kafka was the clear winner among the queuing and event streaming systems. 

Today at Uber, we have one of the largest deployments of Apache Kafka in the industry, with trillions of messages and Petabytes of data per day. As the transport mechanism for sending streaming data to both batch and realtime systems, Kafka at Uber empowers a large number of different workflows, such as propagating event data from the rider and driver apps, enabling a streaming analytics platform (e.g. Apache Samza\cite{samza}, Apache Flink), streaming database changelog to the downstream subscribers, ingesting all sorts of data into Uber’s Apache Hadoop Data Lake. Due to Uber’s large scale, fault-tolerance considerations and some unique requirements, we customized Kafka and added the following enhancements.

\subsubsection{Cluster federation}

To improve the availability and to tolerate single-cluster failure, we at Uber developed a novel federated Kafka cluster setup which hides the cluster details from producers/consumers. Users do not need to know which cluster a topic resides in and the clients view a “logical cluster”. A metadata server aggregates all the metadata information of the clusters and topics in a central place, so that it can transparently route the client’s request to the actual physical cluster. In addition to reliability, cluster federation also improves scalability to support business growth. Based on our empirical data, the ideal cluster size is less than 150 nodes for optimum performance. With federation, the Kafka service can scale horizontally by adding more clusters when a cluster is full. New topics are seamlessly created on the newly added clusters. Lastly, cluster federation also brings the ease of topic management. Inside Uber there are a large number of applications and clients, and it’s challenging to migrate a topic with live consumers between clusters. Typically, this requires manual user coordination to shift their traffic to the new cluster, resulting in job restart. Cluster federation enables consumer traffic redirection to another physical cluster without restarting the application.

\subsubsection{Dead letter queue}

There are cases when some messages fail to be processed by downstream application, for example, due to message corruption or unexpected behavior. In Apache Kafka’s model, there are two options to handle such failed messages: either drop those messages or retry indefinitely which blocks processing of the subsequent messages. However, there are many scenarios in Uber that demand neither data loss nor clogged processing, such as trip receipt processing. To accommodate such use cases, a Dead Letter Queues (DLQ) strategy was built on top of the Kafka interface\cite{dlq}. If a consumer of the topic cannot process a message with several retries, it will publish that message to the dead letter topic. The messages in the dead letter topic can be purged or merged (i.e. retried) on demand by the users. This way, the unprocessed messages remain separate and therefore are unable to impede live traffic.

\subsubsection{Consumer Proxy}

Open source Kafka includes a consumer library which packages sophisticated logic of batching and compression. Though such client-side optimizations improves the consumer throughput, it brings a big challenge to large organizations like Uber regarding client management. With tens of thousands of Kafka applications running, it’s tremendously difficult for the platform team to support the users in troubleshooting and debugging. In addition, it slows down the development of the client library, as it takes months to upgrade the client library in all the applications. Moreover, large organizations use many programming languages, so it’s hard to provide multi-language support when the clients are complex. Lastly, due to Kafka’s architecture limitations, the open-source Kafka limits the number of the instances in a consumer group to no more than the number of the topic’s partition, and therefore puts a cap on consumer’s level of parallelism.

To address these challenges, we built a proxy layer that consumes messages from Kafka and dispatches them to a user-registered gRPC service endpoint for all the pub/sub use cases. The complexities of the consumer library are encapsulated in the proxy layer, and applications only need to adopt a very thin, machine-generated gRPC client. In particular, the consumer proxy provides sophisticated error handling. When the downstream service fails to receive or process some messages, the consumer proxy can retry the dispatch, and send them to the DLQ if several retries failed.

\begin{figure}[h]
  \centering
  \includegraphics[width=\linewidth]{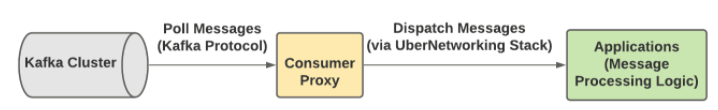}
  \caption{Overview of the Kafka Consumer Proxy at Uber}
  \label{fig:kcp}
\end{figure}

Another noticeable benefit of the consumer proxy is the change of delivery mechanism from message polling to push-based message dispatching, as shown in Figure \ref{fig:kcp}. Most pub/sub use cases inside Uber do not assume any dependencies among the messages. As a result, a push-based dispatching mechanism can greatly improve the consumption throughput by enabling higher parallelism for slow consumers with negligible latency overhead. This addresses Kafka’s consumer group size issue and allows significantly more concurrent processing opportunities to the applications. 

\subsubsection{Cross-cluster replication} \label{sec:kafka-cross-dc}

Given the large-scale use of Kafka within Uber, we ended up using multiple clusters in different data centers. With this setup, the cross-cluster replication of Kafka messages is necessary for two reasons. First, we need to look at the global view of this data for a variety of use cases. For example, in order to compute business metrics related to trips, we need to gather information from all data centers and analyze it in one place. Second, Kafka is also replicated for redundancy to tolerate cluster and datacenter failures. To achieve this, we built and open-sourced a robust and performant replicator across the Kafka clusters called uReplicator\cite{ureplicator}. uReplicator is designed for strong reliability and elasticity. It has an in-built rebalancing algorithm so that it minimizes the number of the affected topic partitions during rebalancing. Moreover, uReplicator is adaptive to the workload so that when there is bursty traffic it can dynamically redistribute the load to the standby workers for elasticity. 

On top of this, to ensure there is no data loss from the cross-cluster replication, we also developed and open sourced an end-to-end auditing service called Chaperone\cite{chaperone}. Chaperone collects key statistics like the number of unique messages in a tumbling time window from every stage of the replication pipeline. The auditing service compares the collected statistics and generates alerts when mismatch is detected.

\bigskip

With these improvements, we built a standardized and reliable streaming and messaging platform on top of Kafka to empower various use cases in real-time. The future work in this area includes scaling up to multi regions across on-prem data centers and cloud, as well as better cost-efficiency architecture, which is discussed in section \ref{sec:future}.

\subsection{Apache Flink for stream processing} \label{sec:flink}

In order to process all the real-time data coming through Kafka, we have built a stream processing platform on top of Apache Flink\cite{katsifodimos2016apache}. Apache Flink is an open-source, distributed stream processing framework with a high-throughput, low-latency engine widely adopted in the industry. We adopted Apache Flink for a number of reasons. First, it is robust enough to continuously support a large number of workloads with the built-in state management and checkpointing features for failure recovery. Second, it is easy to scale and can handle back-pressure efficiently when faced with a massive input Kafka lag. Third, it has a large and active open source community as well as a rich ecosystem of components and toolings. Based on our comparisons done in 2016 with Apache Storm, Apache Spark and Apache Samza, Flink was deemed as the better choice of technology for this layer. Storm performed poorly in handling back pressure when faced with a massive input backlog of millions of messages, taking several hours to recover whereas Flink only took 20 minutes. Spark jobs consumed 5-10 times more memory than a corresponding Flink job for the same workload. Samza had a strict dependency on Kafka for maintaining its internal state, which induced a significant operational overhead.

At Uber, we use Flink heavily for both facilitating customer-facing products and powering internal analytics with a wide range of insights captured from across the world and at all times, such as city-specific market conditions to global financial estimations. The stream processing logic can be expressed by the users in 2 ways, a SQL dialect or a set of low-level APIs. The SQL dialect is commonly used by different categories of users, technical and non-technical, such as engineers, data scientists, operations personnel, product managers and so on. The more advanced users prefer to use the API for expressing complex logic as well as connecting to external systems such as databases, RPC endpoints, caches and so on. To better support Uber use cases, we made the following contributions and improvements to Apache Flink.

\subsubsection{Building streaming analytical applications with SQL}

One of the most important contributions we made was to introduce a layer on top of the Flink framework known as FlinkSQL\cite{athenax}. This is now contributed back to Apache Flink project and it provides the ability to transform an input Apache Calcite\cite{calcite} SQL query into an efficient Flink job. The SQL processor compiles the queries to reliable, efficient, distributed Flink applications, and manages the full lifecycle of the application, allowing users to focus solely on their business logic. Internally, it converts the input SQL query into a logical plan, runs it through the query optimizer and creates a physical plan which can be translated into a Flink job using the corresponding Flink API. As a result, users of all technical levels can run their streaming processing applications in production in a span of mere hours regardless of scale.

These internal details are hidden from the user which has a big trade-off. It makes adoption very easy since all the users need to understand is the data source, e.g. input Kafka topic, and the SQL syntax. However, it adds significant operational overhead for the platform team to tune and maintain the production jobs. In particular, we had to overcome the following challenges:

{\bfseries Resource estimation and auto scaling} The resource configurations such as allocated CPU and memory are important for the job health and also the cost efficiency. We used empirical analysis for establishing a correlation between the common job types and the corresponding resource requirements. For instance, a stateless Flink job which does not maintain any aggregation windows is CPU bound vs a stream-stream join job will almost always be memory bound. We also observed that the job load may vary during peak and off-peak hours. To maximize the cluster utilization, we employ continuous monitoring of the job load and garbage collection statistics, and perform auto-scaling when necessary.

{\bfseries Job monitoring and automatic failure recovery} Since the end user does not know about the underlying Flink job and its job status, the platform needs to monitor the job and provide a strong reliability guarantee. To address this, we built a component for automatically handling job failures when it detects a certain condition. It is a rule-based engine which compares the Flink job’s key metrics such as resource usage against the desired state and takes corrective action such as restarting a stuck job or auto scaling.

Note that FlinkSQL has different semantics from the batch processing SQL systems such as Presto. FlinkSQL is a stream processing engine wherein both the input and output are unbounded streams. Whereas, batch processing engines query bounded datasets and output a bounded dataset. One of the future work for FlinkSQL is to unify the streaming/batch processing semantics as discussed in section \ref{sec:future}.

\subsubsection{Unified architecture for deployment, management and operation} \label{sec:unified-flink-architecture}

\begin{figure}[h]
  \centering
  \includegraphics[width=\linewidth]{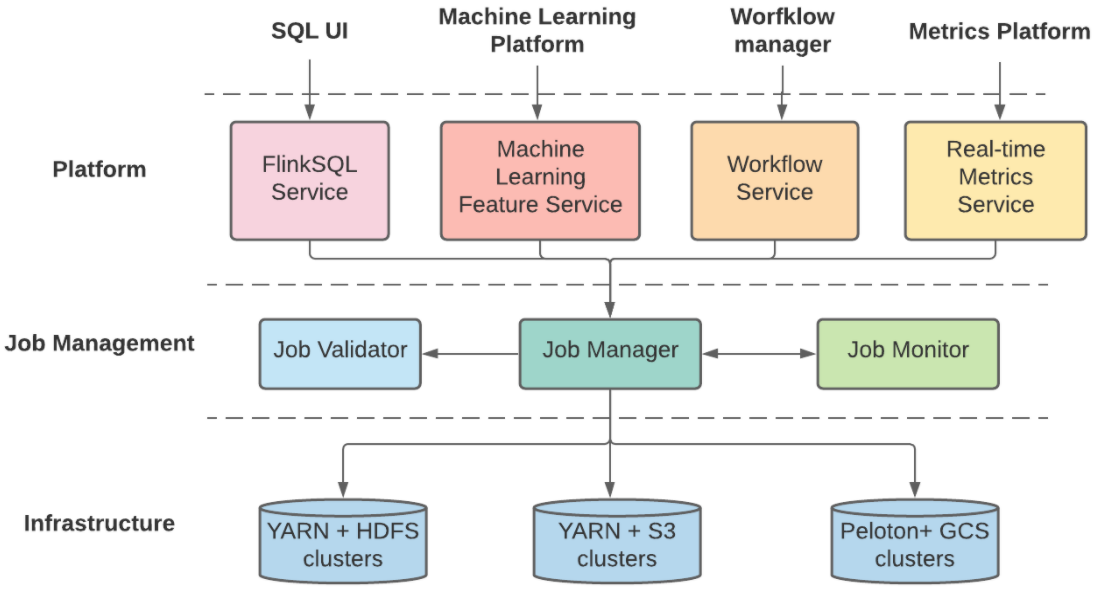}
  \caption{The layers of the Unified Flink architecture at Uber}
  \label{fig:flink}
\end{figure}

Since we have provided two platforms to the users for building and managing the stream processing pipelines, we identified commonalities between the two and converged them into a unified architecture for deployment, management and operation. The new unified platform also addressed several other challenges and resulted in a layered architecture for better extensibility and scalability as depicted in Figure \ref{fig:flink}. 

The Platform layer handles organizing the business logic and integration with other external platforms such as Machine learning feature training, workflow management and SQL compilation. It consists of multiple business-specific components and can easily be extended to support new business requirements with additional components. It transforms the specific business logic into standard Flink job definitions, and passes this to the next layer for validation and management.

The job management layer manages the Flink job's lifecycle including validation, deployment, monitoring and failure recovery. It offers a set of unified API abstractions for the platform layer (such as Start/Stop/List a job) and persists the job information including the state checkpoints and the metadata. In addition, it serves as the proxy layer to the physical clusters and dispatches the jobs based on the job type, importance and priority. Lastly, a shared component in the job management server continuously monitors the health of all jobs and automatically recovers the jobs from the transient failures.

The bottom layer is the infrastructure layer consisting of the compute clusters and storage backend. It also provides the abstraction of the physical resources for flexibility and extensibility, regardless of the hosting infrastructure being on-prem or cloud. For example, the compute clusters can be paired with different resource schedulers such as YARN and Uber’s Peloton\cite{peloton}. Similarly, the storage backend can adopt HDFS, Amazon S3\cite{s3} or Google Cloud Storage\cite{gcs} for the state checkpoints, to meet the various requirements for storage choice. 

With these improvements, Flink has emerged as the de facto stream processing platform within Uber, powering several thousands of jobs with 30\% year-over-year growth rate. Based on user feedback, the current challenges in this layer include the need for seamless data backfills without writing additional code which is discussed in detail in Section \ref{sec:backfill}. Furthermore, some critical use cases need the ability to do a Flink job restart without any downtime which is an active area of investigation in the Flink community.

\subsection{Apache Pinot for OLAP} \label{sec:pinot}

Apache Pinot\cite{im2018pinot} is an open-source, distributed, OnLine Analytical Processing (OLAP) system designed for performing low latency analytical queries on terabytes-scale data. Pinot employs the lambda architecture to present a federated view between real-time and historical (offline) data. As a column store, Pinot supports a number of fast indexing techniques, such as inverted, range, sorted and startree index\cite{im2018pinot}, to answer the low-latency OLAP queries. Pinot takes a scatter-gather-merge approach to query large tables in a distributed fashion: data is chunked by time boundary and grouped into segments; while the query is first decomposed into sub-plans which execute on the distributed segments in parallel, and then the plan results are aggregated and merged into a final one. 

We decided to adopt Apache Pinot as our OLAP solution for several reasons. During this time in 2018, the only other options available were Elasticsearch\cite{elastic} and Apache Druid\cite{yang2014druid}. Based on our experimental evaluation and outside research, Apache Pinot has a smaller memory and disk footprint as well as supports significantly lower query latency SLA. In particular: 
\begin{itemize}
\item ElasticSearch: With the same amount of data ingested into Elasticsearch and Pinot, Elasitcsearch's memory usage was 4x higher and disk usage was 8x higher than Pinot. In addition, Elasticsearch's query latency was 2x-4x higher than Pinot, benchmarked with a combination of filters, aggregation and group by/order by queries.

\item Apache Druid: Pinot is similar in architecture to Apache Druid but has incorporated optimized data structures such as bit compressed forward indices, for lowering the data footprint. It also uses specialized indices for faster query execution such as Startree \cite{im2018pinot}, sorted and range indices, which could result in order of magnitude difference of query latency. Recent studies done outside of Uber also confirmed this performance benefit of Pinot over Druid \cite{compareOlap} \cite{confluera}. 

\end{itemize}

At Uber, Pinot powers a number of real-time analytics use cases. Various products build their customized dashboards on Pinot for visualizing and exploring important metrics such as rides demand-supply, UberEats orders statistics and so on. Another category of use cases stems from the need to execute analytical queries as part of many backend services. The primary distinguishing requirement for such use cases is data freshness and query latency which needs to be real-time in nature. For example, identifying rider cancellation or abandoned UberEats carts instantly, enables quick corrective action in the form of messaging and incentives. We have contributed the following enhancements to Apache Pinot to handle Uber’s unique requirements around high availability, rich query support and exactly once semantics (i.e. upserts).

\subsubsection{Upsert support}

Upsert is a common requirement by many use cases inside Uber, such as correcting a ride fare and updating a delivery status. We designed and developed a scalable upsert solution on Pinot, so that records can be updated during the real-time ingestion into the OLAP store. To the best of our knowledge, Apache Pinot is the only open-source real-time OLAP store that supports upsert. The key technical challenge for upsert is tracking the locations of the records with the same primary key. In a real-time system, it’s very complicated and inefficient to keep track of these locations in a centralized manner and coordinate with distributed storage nodes. To overcome this challenge, we organize the input stream into multiple partitions by the primary key, and distribute each partition to a node for processing. As a result, all the records with the same primary key are assigned to the same node. On top of that, we introduced a new routing strategy that dispatches subqueries over the segments of the same partition to the same node to ensure the integrity of the query result. Together they lead to a shared-nothing solution to this problem in Pinot. This shared-nothing solution has many advantages, including better scalability, elimination of single point of failure, and ease of operation.

\subsubsection{Full SQL support }

Pinot is an OLAP system that excels at low-latency queries with a rich set of indexing technique support. However, it lacks several notable SQL features such as subqueries and joins. To fill this gap, we have integrated Pinot with Presto for enabling standard PrestoSQL queries on Pinot tables\cite{pinot-sql}. In fact, Presto is the defacto query engine for interactive queries within Uber. This combination works great since we can combine Pinot’s seconds level data freshness with Presto’s flexibility in doing complicated queries. In addition, predicate pushdowns and aggregation function pushdowns enable us to achieve sub-second query latencies for such PrestoSQL queries - which is not possible to do on standard backends such as HDFS/Hive. 

\subsubsection{Integration with the rest of Data ecosystem}

In large corporations like Uber, it’s a priority to improve engineering productivity and development velocity, in an environment where every product evolves at a fast pace. Towards this goal, we’ve spent a lot of time integrating Pinot with the rest of the Data ecosystem to ensure a seamless user experience\cite{operate-pinot}. Pinot integrates with Uber’s schema service to automatically infer the schema from the input Kafka topic and estimate the cardinality by sampling the messages. Pinot also integrates with FlinkSQL as a data sink, so customers can simply build a SQL transformation query and the output messages can be “pushed” to Pinot. Similar integrations have been added to Piper, Uber’s data workflow management system\cite{piper}, to create Pinot offline tables from Hive datasets via Spark.

\subsubsection{Peer-to-peer segment recovery}

The original design of Apache Pinot introduced a strict dependency on an external archival or “segment store” such as HDFS, Amazon S3, Google GCS and so on. During real time data ingestion, completed segments had to be synchronously backed up to this segment store to recover from any subsequent failures. In addition, this backup was done through one single controller. Needless to say, this was a huge scalability bottleneck and caused data freshness violation. Moreover, any segment store failures caused all data ingestion to come to a halt. Our team designed and implemented an asynchronous solution wherein server replicas can serve the archived segments in case of failures. Thus, we replaced a centralized segment store with a peer-to-peer scheme, while still maintaining the same guarantees around data and query consistency. Lastly, this also solved the single node backup bottleneck and significantly improved overall data freshness. 

\bigskip

With these improvements, Pinot adoption has grown significantly within Uber. In 2 years since it was introduced in our Data stack, the data footprint has grown from dozens of GBs of data to over several hundreds of TBs of data. At the same time the query workload has grown from a few hundreds of QPS (Queries Per Second) to tens of thousands of QPS. Our team continues to address the growing needs within Uber and are currently working on the following challenges:

{\bfseries Ability to perform low latency joins}: Currently joins are performed by Presto, which federates query execution across Pinot and Hive. However, this is done entirely in-memory in the Presto worker and cannot be used for critical use cases. We are contributing the ability to perform lookup joins to Pinot to support joining tables with commonly used dimension tables.

{\bfseries Semistructured (e.g. JSON) data support}: Users currently rely on a Flink job to preprocess an input Kafka topic with nested JSON format into a flattened-schema Kafka topic for Pinot ingestion. We are working with the community in building native JSON support for both ingestion and queries.

\subsection{HDFS for archival store }

In Uber, we use HDFS as the long term storage for all the Data. Most of this data comes from Kafka which is in Avro format and is persisted in HDFS as raw logs. These logs are then merged into the long term Parquet data format using a compaction process and made available via standard processing engines such as Hive, Presto or Spark. Such datasets constitute the source of truth for all analytical data. This is used to backfill data in Kafka, Pinot and even some OLTP or key-value store data sinks. In addition, HDFS is used by other platforms for managing their own persistent storage. For instance, Apache Flink uses HDFS for maintaining the job checkpoints. This consists of all the input stream offsets as well as snapshots of the Flink job's internal state per container. Furthermore, Apache Pinot uses HDFS for long term segment archival which is crucial for correcting failed replicas or during server bootstrap. 

\subsection{Presto for Interactive Query}

Traditionally, in the big data ecosystem a distributed SQL query engine such as Hive\cite{hive} is used for processing batch datasets where the emphasis is on query flexibility rather than ingestion or query latency. In the recent years, there have been increasing demands on the interactive analytics workloads to derive the insights in a quick manner and at Uber we adopted Presto\cite{presto} as the interactive query engine solution. Presto is an open source, distributed query engine originally developed by Facebook. Presto was designed from the ground up for fast analytical queries against large scale datasets by employing a Massively Parallel Processing (MPP) engine and performing all computations in-memory, thus avoiding materialization overhead of writing intermediate results to disk. 

Moreover, Presto is designed to be flexible and extensible. It provides a Connector API with high performance I/O interface to multiple data sources, including Hadoop data warehouses, RDBMSs and NoSQL systems. In the case of Uber, data scientists and engineers often want to do exploration on real-time data to enhance the sensitivity of the corresponding features or models. In order to achieve this, we have leveraged Presto’s connector model and built a Pinot connector to deeply integrate with Apache Pinot so that we can execute standard Presto SQL queries on fresh data. One challenge we overcame during this connector development is to be intelligent and selective on which parts of the physical plan can be pushed down to the Pinot layer. Our first version of this connector only included predicate pushdown given the limited connector API. In order to lower query latency and leverage Pinot’s fast indexing, we enhanced Presto’s query planner and extended Presto Connector API to push as many operators down to the Pinot layer as possible, such as projection, aggregation and limit.

\section{Use cases analysis} \label{sec:use-case}

In this section, we present several different real-time use cases across the 4 broad categories (as in Figure \ref{fig:abstractions}) in production at Uber, and show how they use the different systems to achieve their business goals. Also, we discuss the design tradeoffs considered by those use cases.

\subsection{Analytical Application: Surge Pricing}

The surge\cite{garg2019driver} use case is a dynamic pricing mechanism in Uber ride-hailing marketplace to balance the supply of available drivers with the demand for rides. On the rider side, surge pricing reduces the demand to match the level of available drivers and maintains the reliability of the marketplace. On the driver side, it encourages drivers to drive during certain hours and locations, as drivers earn more during surge. 

Surge pricing is essentially a streaming pipeline for computing the pricing multipliers per hexagon-area geofence based on the trip data, rider and driver status in a time window. The surge pricing pipeline ingests streaming data from Kafka, runs a complex machine-learning based algorithm in Flink, and stores the result in a sink key-value store for quick result look up. The surge pricing favors data freshness and availability over data consistency. The late-arriving messages do not contribute to the surge computation and the pipeline must meet a strict end-to-end latency SLA requirement on the calculation per time window. This tradeoff is reflected in the design that the surge pricing pipeline uses the Kafka cluster configured for higher throughput but not lossless guarantee, as well as an active-active setup for higher availability which is described in Section \ref{sec:all-active}.

\subsection{Dashboards: UberEats Restaurant Manager}
Dashboards are popular for observing trends and spotting anomalies at a glance. And at Uber, many engineering teams build customized dashboards using the real-time analytics systems. Among them, UberEats restaurant manager is a good representative example. This dashboard enables the owner of a Restaurant to get insights from the UberEats orders regarding customer satisfaction, popular menu items, sales and service quality analysis, via generated interactive, slice-and-dice queries. 

At a high level, the restaurant manager demands fresher data and low query latency, but does not require too much flexibility as the patterns of the generated queries are fixed. To meet the requirements, we used Pinot with the efficient pre-aggregation indices of the large volume of raw records, in order to reduce the serving time. Also, we built preprocessors in Flink such as aggressive filtering, partial aggregate and roll-ups to further reduce the processing time in Pinot and meet the latency SLA. With such preprocessing, we trade the query flexibility required for ad-hoc exploration and complexity of query evolution for lower latency. 

In general, there is a tradeoff between processing at the transformation time, as done by Flink, and processing at query time, as done by Pinot. The preprocessing during transformation time can create optimized indices and reduce the amount of data for serving, but it reduces the query flexibility on the serving layer.

\subsection{Machine Learning: Real-time Prediction Monitoring} \label{sec:ml-monitoring}

Machine learning (ML) has been playing a crucial role within Uber to create seamless, impactful experiences for our customers\cite{michelangelo}. To ensure ML model quality, it is critical to monitor its predictions so as to ensure that the data pipelines are continuing to send accurate data. To address this, a real-time prediction monitoring pipeline is set up that joins the predictions to the observed outcomes (or labels) generated by the data pipeline, creating ongoing, live measurements of model accuracy.

The key requirement from this use case is scalability, due to a high volume and high cardinality of data to be processed. With thousands of ML models deployed and each model with hundreds of features, there are several hundreds of thousands of time series with millions of data points computed per second, far beyond the capability of the time-series database inside Uber. Thanks to the horizontal scalability of Flink, we deployed a large streaming job to aggregate the metrics and detect prediction abnormality. To boost the query performance over the large number of data points, the Flink job also creates pre-aggregation as Pinot tables.

This real-time prediction monitoring pipeline represents a large number of use cases that build real-time OLAP cubes with pre-aggregates and indices in Pinot, to speed up query execution time and throughput for large scale datasets.

\begin{table}
  \caption{The components used by the example use cases}
  \label{tab:components}
  \begin{tabular}{ m{4em} | m{0.8cm} | m{1.3cm} | m{1.6cm} | m{1.3cm} }
    \toprule
    & Surge  & Restaurant Manager  & Real-time Prediction Monitoring & Eats Ops Automation\\
    \midrule
    API  & Y &   & Y &   \\ 
     SQL  &   & Y & Y & Y \\ 
     OLAP &   & Y & Y & Y \\ 
     Compute & Y & Y & Y & Y \\ 
     Stream  & Y & Y & Y & Y \\ 
     Storage  &  & Y & Y &  \\ 
  \bottomrule
\end{tabular}
\end{table}


\subsection{Ad-hoc Exploration: UberEats Ops Automation}

The UberEats team needed a way to execute ad hoc analytical queries on real time data generated by couriers, restaurants and eaters. Once an insight was discovered, a subsequent need was to productionize the query in a rule-based automation framework. This was a critical component used by the Ops team to combat Covid 19 and keep restaurants open in different geographical locations like Europe. To comply with regulation and safety rules, Uber needed to limit the number of customers and couriers at a restaurant. The ops team was able to identify such metrics using Presto on top of real-time data managed by Pinot and then inject such queries into the automation framework. 

This framework uses Pinot to aggregate needed statistics for a given geographical location in the past few minutes and then generates alerts and notifications to the couriers and restaurants accordingly. Thus the same infrastructure provided a seamless path from ad-hoc exploration to production rollout. Needless to say, the underlying system has to be extremely reliable and scalable since this decision making process is critical not only to the business but also for the safety of the customers. Pinot, Presto and Flink were able to scale easily with the organic data growth and performed reliably during peak hours. 

\bigskip

To summarize, Table ~\ref{tab:components} shows the components in the real-time infrastructure used by the representative use case for each category.

\section{All-active strategy} \label{sec:all-active}

Providing business resilience and continuity is a top priority for Uber. Disaster recovery plans are built carefully to minimize the business impact from natural and man-made disasters, such as power outages, catastrophic software failures and network outages. At Uber, we rely on a multi-region strategy that ensures services are deployed with backup in data centers geographically distributed, and when the physical infrastructure in one region is unavailable, the service can still stay up and running from other regions.

The foundation of this multi-region real-time architecture is a multi-region Kafka setup that provides data redundancy and traffic continuation support for its clients. In fact, the majority of the services in the stack above depend on Kafka for the active/active setup. For example, Figure \ref{fig:active-active} below shows how Uber’s dynamic pricing service (i.e. surge pricing) uses active-active Kafka to build the disaster recovery plan. All the trip events are sent over to the Kafka regional cluster and then aggregated into the aggregate clusters for the global view. Then in each region a complex Flink job with large-memory footprint will compute the pricing for different areas. Each region has an instance of ‘update service’ and one of them is labelled as primary by an all-active coordinating service. The update service from the primary region stores the pricing result in an active/active database for quick lookup.

\begin{figure}[h]
  \centering
  \includegraphics[width=\linewidth]{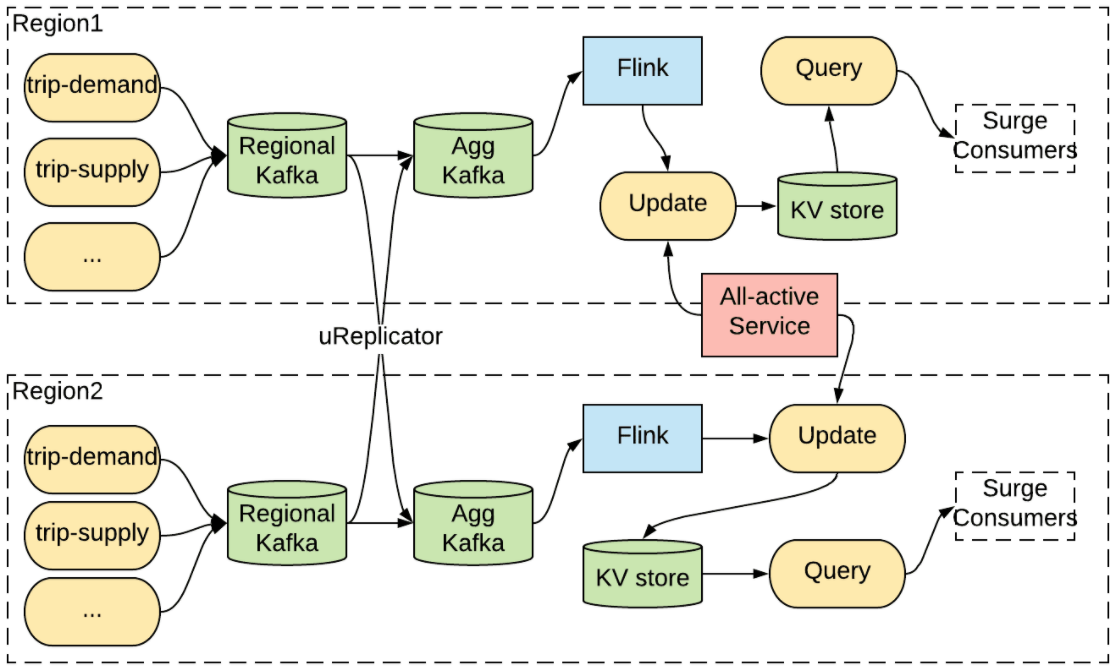}
  \caption{The active-active setup for surge pricing}
  \label{fig:active-active}
\end{figure}

When disaster strikes the primary region, the active-active service assigns another region to be the primary, and the surge pricing calculation fails over to another region. It’s important to note that the computation state of the Flink job is too large to be synchronously replicated between regions, and therefore its state must be computed independently from the input messages from the aggregate clusters. Given that the input to the Flink job from aggregate Kafka is consistent across all regions, the output state converges. This approach is compute intensive since we’re running redundant pipelines in each region.

The other strategy is to consume Kafka in an active/passive mode: only one consumer (identified by a unique name) is allowed to consume from the aggregate clusters in one of the regions designated as the primary region at a time. When disaster happens, the service can fail over to another region and resume its consumption progress. Such active/passive mode is desirable for the services that favor strong consistency such as payment processing and auditing. 

\begin{figure}[h]
  \centering
  \includegraphics[width=\linewidth]{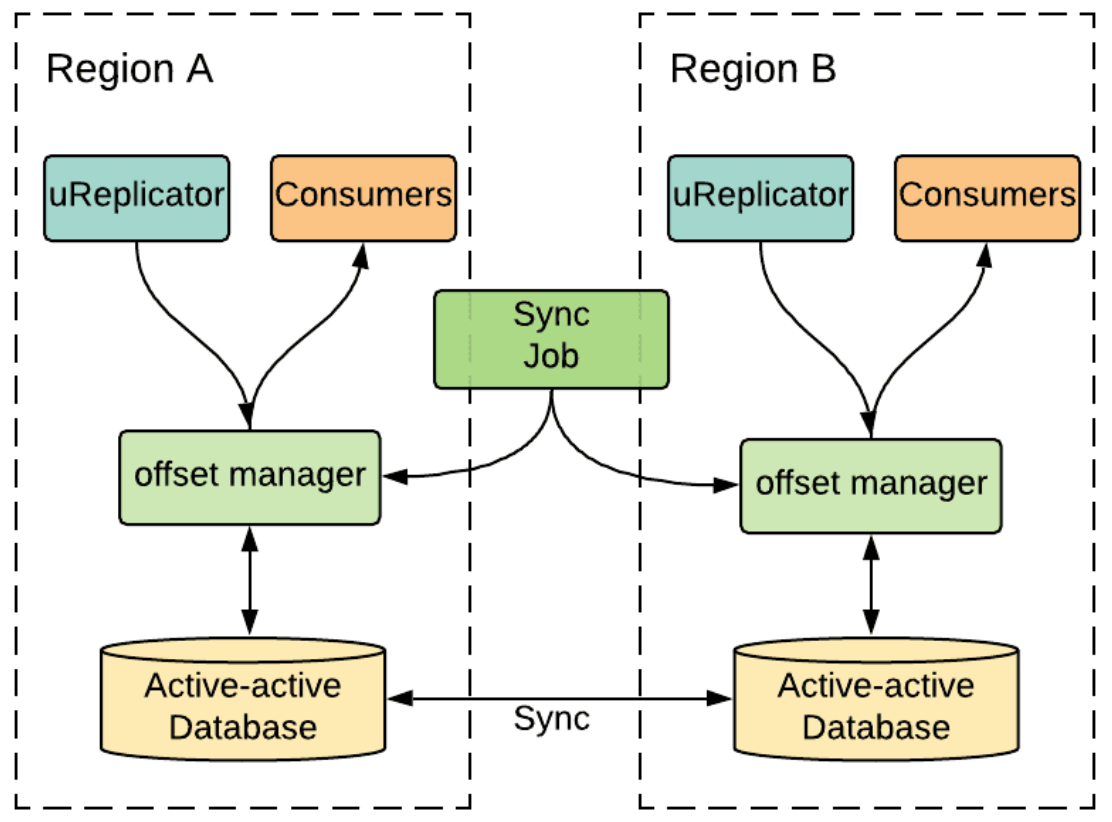}
  \caption{The active-passive setup for stronger consistency}
  \label{fig:active-passive}
\end{figure}

As the consumption progress is represented by the offset of the Kafka topic, the key challenge of the active/passive strategy is offset synchronization of consumers across regions. Because many services at Uber cannot accept any data loss, in case of a failover, the consumer can neither resume from the high watermark (i.e. the latest messages), nor from the low watermark (i.e. the earliest messages) to avoid too much backlog. In order to overcome the challenge of offset mappings across the regions, we developed a sophisticated offset management service at Uber. As shown in Figure \ref{fig:active-passive}, when uReplicator (introduced in section \ref{sec:kafka-cross-dc}) replicates messages from source cluster to the destination cluster, it periodically checkpoints the offset mapping from source to destination in an active-active database. Meanwhile, an offset sync job periodically synchronizes the offsets between the two regions for the active-passive consumers. So when an active/passive consumer fails over from one region to another, the consumer can take the latest synchronized offset and resume the consumption.

\section{Backfill} \label{sec:backfill}

There is a recurring need to go back in time and reprocess the stream data at Uber, for several reasons. First, a new data process pipeline often needs to test against the existing data, or a new machine learning model often needs to be trained with a few months of data. To save time and be able to iterate faster, the testing or training is done on historic data that is already available. Second, sometimes a bug may be discovered in a real-time application that has already processed the data for a period. In such cases there could be a desire to reprocess some/all of the data separately after fixing the bug. Third, similar to the previous case, there can be a change of stream processing logic that necessitates reprocessing of old data.

The backfill problem appears to be a common problem wherever there is realtime big data processing. Lambda\cite{lambda} and Kappa\cite{kappa} architectures have been proposed in this respect but both suffer from limitations. Lambda architecture maintains two separate systems: one for batch, and one for stream processing. This leads to maintenance and consistency issues when trying to keep both implementations in sync. Kappa architecture improves upon this by using the same streaming code for both real-time and backfill processing but requires very long data retention in Kafka and may not be very efficient in terms of processing throughput. Given the scale of data generated into Kafka at Uber and the operational concerns regarding node replacement, we limit Kafka retention to only a few days. Therefore, we're unable to adopt the Kappa architecture.

At Uber, we built a solution for ease of backfill for stream processing use cases using Flink which has 2 modes of operations:

\begin{itemize}
    \item SQL based: We added the ability to execute the same SQL query on both real-time (Kafka) and offline datasets (Hive). In this case, the FlinkSQL compiler will translate the SQL query to two different Flink jobs: one using DataStream API and the other using DataSet API. Although this is similar to Lambda architecture, the user does not need to maintain 2 distinct jobs.
    
    \item API based: This solution is internally named as Kappa+\cite{kappa}. The Kappa+ architecture is able to reuse the stream processing logic just like Kappa architecture but it can directly read archived data from offline datasets such as Hive. The Kappa+ architecture addressed several issues on processing the batch datasets with streaming logic, such as identifying the start/end boundary of the bounded input, handling the higher throughput from the historic data with throttling, fine tuning job memory as the offline data could be out of order and therefore demand larger window for buffering. Effectively, using Kappa+ we can execute the same code with minor config changes on both streaming or batch data sources.
\end{itemize}

This is an active area of investigation and there are lots of edge cases that need to be handled in both these solutions. A full evaluation of each approach is out of scope of this paper.

\section{Related Work} \label{sec:related}

Real-time data infrastructure spans a wide range of components, and there are plentiful related systems in each area.

{\bfseries Messaging systems.} Traditional enterprise messaging systems ActiveMQ\cite{activemq} RabbitMQ\cite{rabbitmq} Oracle Enterprise Messaging Service\cite{oraclems} IBM Websphere MQ\cite{ibmsq} have existed for a long time and often play a critical role as an event bus for processing asynchronous data flows. However, none of those are comparable to Kafka on features, ecosystem and system performance. Recently, a new messaging system Apache Pulsar\cite{pulsar} emerged with a novel tiered architecture\cite{ramasamy2019unifying} that decouples data serving and data storage for better elasticity and easier operation. However, Pulsar is still relatively new and not as mature as Kafka. 

{\bfseries Stream Processing Systems.} The need for highly-scalable stream processing systems has led to the creation of a number of systems in the recent years, including both open-source software like Storm\cite{storm}, Samza\cite{samza}, Heron\cite{kulkarni2015twitter}, Spark Streaming\cite{zaharia2013discretized}, Apex\cite{apex}, and the home-grown ones from large internet companies like Google's Photon\cite{ananthanarayanan2013photon}, Facebook’s Puma/Swift\cite{chen2016realtime}, Amazon's Kinesis\cite{kinesis}. In addition to overcoming the challenges of scalability, efficiency and fault tolerance, another important ongoing trend for these systems is the unification of streaming and batch processing. Systems like Apache Flink\cite{carbone2015apache} are expanding its architecture to support batch processing use cases, while frameworks like Dataflow\cite{akidau2015dataflow} and Apache Beam\cite{beam} approach this via an abstraction layer over different processing engines.

{\bfseries Real-time OLAP Systems.} Real-time OLAP systems have become popular in recent years, as modern businesses need to quickly transform freshly obtained data into insights. Apache Druid\cite{yang2014druid} and Clickhouse\cite{clickhouse} are the open source systems commonly adopted in the industry. They are like Pinot to buffer ingested streams and utilize column stores to achieve efficient column scans. Helios\cite{potharaju2020helios} is a similar real-time OLAP store developed and used at Microsoft. Another way to improve the performance of the OLAP system is to pre-aggregate data into cubes, then perform query execution on the pre-aggregated data \cite{kylin}. However, such performance improvements come at the expense of query inflexibility. HTAP databases are another category of emerging systems to unify transactional and analytical processing in a single system \cite{kemper2011hyper} \cite{huang2020tidb} \cite{farber2012sap}. However, one challenge is to have a clean separation of the two to prevent the interference of the analytical queries over the operational workload\cite{psaroudakis2014scaling} despite some recent attempts on this problem\cite{makreshanski2017batchdb}.
 
{\bfseries SQL systems.} Systems that run SQL against large data sets have become popular over the past decade. Each of these systems present a unique set of tradeoffs, and a comprehensive examination of the space is outside the scope of this paper. Apache Hive\cite{hive} was originally developed at Facebook to provide a SQL-like interface over data stored in HDFS\cite{hdfs}, and Dremel\cite{melnik2020dremel} is an exa-scale columnar SQL query engine that is optimized for large complex ad-hoc queries used at Google. Spark SQL\cite{armbrust2015spark} is a more modern system built on the popular Spark engine addressing many of the limitations of MapReduce\cite{dean2008mapreduce}. Systems like MySQL\cite{mysql}, Impala\cite{impala}, Drill\cite{drill} are common open source databases that can be used for analytical purposes. In recent years, more SQL systems extended the support of querying real-time data. Procella\cite{chattopadhyay2019procella} is a highly scalable and performant SQL engine used in YouTube with native support for lambda architecture and low-latency data ingestion. Trill\cite{chandramouli2014trill} is a query processor from Microsoft that handles streaming and relational queries with early results, across the latency spectrum from real-time to offline. We chose the open source Presto for its interactiveness, flexibility and extensibility that makes it easy to integrate with other data sources and databases via the Connector API, and enhanced it with the real-time data availability via Pinot.

Real-time data powers many use cases at other very large scale internet companies. Chen et al\cite{chen2016realtime} presented the realtime data processing and analytics infrastructure at Facebook. It shares a similar full-stack architecture with ours, towards a similar scale and targeting latency of seconds. Most components in their architecture were developed in house and private, while at Uber we tried to adopt open source solutions and leverage the wider community. F1 Lightning\cite{yang2020f1} is a HTAP solution from Google that provides analytical processing over change data streamed from transactional stores, and highlights a federated query engine loosely coupled with multiple transaction data stores such as F1 DB\cite{shute2012f1} and Spanner\cite{corbett2013spanner}. The real-time data infrastructure we built at Uber not only integrates with the transactional data via Change Data Capture (CDC), but also works directly over natively generated streaming data. Recently, hybrid serving and analytical processing (HSAP) has emerged as a new kind of architecture that fuses analytical processing and serving as well as the online and offline analysis. Alibaba’s Hologres\cite{jiang2020alibaba} is an example of such architecture that powers Alibaba’s internal big data stack as well as its public cloud offerings. Though such hybrid architecture can lead to more efficiency and enable more optimization, at Uber we chose employing loosely coupled independent systems for the ease of customization and evolution of each component.

\section{Lessons learned} \label{sec:lessons}

We have learned many lessons in our journey of building and scaling the real-time data infrastructure at Uber.

\subsection{Open source adoption}

As seen before, most of the real-time analytics stack and in fact the larger data stack in Uber has been built on open source technologies. The primary reason behind this philosophy is the need to iterate quickly. The engineering needs at Uber are constantly evolving and the ability to deliver a quick solution is crucial. Relying on open source gives us a strong foundation to build upon and reduces the time to market. Naturally, this also helps in handling churn in a graceful way. 

However, this is not without its challenges. In our experience, most open source technologies were built for a specific purpose and at Uber, we had to make it work for a lot of dimensions such as a wide spectrum of use cases, programming language, Uber’s underlying infrastructure, security aspects and so on. For instance, Apache Kafka (circa 2014) was meant to be used primarily for log propagation and used with Java applications. We had to build a RESTful ecosystem around Kafka to make it work with 4 languages: Java, Python, Go and NodeJS which was in use at Uber. This also meant we had to invest in building our own throttling mechanism, metadata discovery, client side failure handling and so on. In addition, we customized the core routing and replication mechanism to handle specialized use cases such as zero data loss for financial data, dead letter queue on top of Kafka. Other such examples of customization include integrating with Uber’s container ecosystem, security policies, building a full SQL layer on top of Apache Pinot for our non engineering audience and seamless backfill using Apache Flink.

\subsection{Rapid system development and evolution}

For a large company like Uber, it’s common to see multiple driving forces to the architecture evolution, such as new business requirements, industrial trends, regulation and compliance, and growth. As a result, one lesson we learned is on the importance of enabling rapid software development so that each system can evolve quickly and independently. 

On the client side, it’s important to set up best practices to manage the large fleet of applications. First, interface standardization is critical so that a clean boundary is established between the services to minimize the risks of breaking the clients. At Uber, we leverage Monorepo\cite{monorepo} to manage all projects in a single code repository, in order to review the changes by the stakeholders and detect the issues early. Second, a thin client is always preferred in order to reduce the frequency of the client upgrades. For example, upgrading Kafka clients used to take several months prior to the introduction of a RESTful, thin Kafka client. Third, language consolidation is another strategy we employ to reduce the number of clients and ways to interact with the system. For low-level programming languages, we purposely reduced the support to only two languages Java and Golang; and for high-level SQL languages, we chose PrestoSQL as the common language for the majority of the use cases, and built connectors to other databases (e.g. Pinot, MySQL).

On the server side, we integrated all our infrastructure components with Uber’s proprietary CI/CD (Continuous Integration/ Continuous Deployment) framework. This ensures that open source software updates as well as internal feature additions are continuously tested and deployed in a staging environment. This also enables continuous end-to-end testing for the mission critical applications and minimizes any production issues. 

\subsection{Ease of operation and monitoring}

Scaling the infrastructure is always a challenge. With rapid business growth, the engineering teams constantly revisit capacity and add more nodes, clusters and data centers. Typically, the speed of scaling physical infrastructure is much faster than scaling the engineering team. As a consequence, lots of manual operations today must be automated in order to sustain the business growth. In fact, at Uber we strategically invested in automation and built declarative frameworks to orchestrate the system deployments. System operators express high-level intentions on operations like cluster turn up and down, resource reallocation, or traffic rebalancing, and the frameworks carry out the instructions without engineer intervention via techniques like configuration generation, containerization and predefined maintenance workflows.

Real-time monitoring and alerting is critical for system reliability and minimizing negative business impact. In addition to cluster wide monitoring, we also provide automated dashboards, alerts and chargeback mechanisms for each use case pertaining to Kafka, Flink or Pinot. This enables the use case owner to monitor health as well as optimize resource utilization.

\subsection{Ease of user onboarding and debugging}

Given the small amount of engineering teams maintaining the underlying technologies, it’s important to build a self-serve system that automates most of the user onboarding, failure handling and triaging for our users. With this in mind, we invested in efforts around the following areas to overcome the challenge of scaling users:

{\bfseries Data discovery.} We use a centralized metadata repository within Uber which is the source of truth for schemas across both realtime and offline systems such as Kafka, Pinot and Hive. This makes it very convenient for users to discover the required datasets. In addition, this system also tracks the data lineage representing flow of data across these components.

{\bfseries Data auditing.} Business events generated by applications are constantly being audited in micro batches from the source all the way to archival. Each such event is decorated with additional metadata such as a unique identifier, application timestamp, service name, tier by the Kafka client. As the events flow from Kafka (regional, aggregate) to Flink or Pinot or Hive, this metadata is used for tracking data loss, duplication for every stage of this data ecosystem as described in Section \ref{sec:kafka-cross-dc}. This makes it very easy for users to detect issues across all of Uber’s data centers. 

{\bfseries Seamless onboarding.} Kafka topics used for application logs are automatically provisioned when the corresponding service is deployed in the production environment. These topics are also automatically expanded as the usage increases along with quota enforcement for limiting the maximum capacity. In a similar vein, users can automatically create Flink and Pinot pipelines using a convenient drag and drop UI that hides the complex sequence of provisioning and capacity allocation\cite{uworc}.

\section{Conclusion} \label{sec:conclusion}

As seen in this paper, the real-time data infrastructure has proliferated at Uber, and the whole stack is powering a lot of mission-critical use cases within Uber. This stack has been optimized for flexibility and scale for different user categories and has been running reliably in production for several years, processing multiple petabytes of data per day. The adoption of open source technologies saved a lot of engineering cost and drastically reduced time to market for our analytical products. The unique contributions by Uber’s engineering teams to all these technologies helped overcome the 3 fundamental scaling challenges, which is summarized below:

{\bfseries Scaling Data} Introduction of Kafka thin client libraries, cluster federation and other techniques discussed above have enabled seamless adoption of Kafka by every service in Uber, making it one of the largest deployments in the world. It provides a robust foundation for orchestrating Flink and Pinot data pipelines that are being leveraged for mission critical use cases across all business units. Flink job automation in terms of deployment and failure recovery has promoted widespread adoption with low operational overhead. We were also able to overcome the lack of high availability SLA for our data archival (HDFS) layer with the investments in Flink’s robust checkpoints and Pinot’s peer-to-peer segment recovery scheme.

{\bfseries Scaling use cases} We invested heavily in flexibility of use of individual technologies for powering varied use cases described above. For instance, with the same client protocol (Apache Kafka consumer) we’re able to serve a wide spectrum of use cases from logging which trades off data consistency for achieving high availability, to disseminating financial data that needs zero data loss guarantees in a multi region ecosystem. Similarly, Pinot provides a low latency OLAP layer for mission-critical use cases as well as enables real-time data exploration via Presto integration. Each such technology can be finely tuned depending on the exact set of requirements. 

{\bfseries Scaling users} Finally, we were able to add a layer of indirection between our users and the underlying technologies using abstractions and standard interfaces, greatly reducing the user support cost. For instance, the introduction of the FlinkSQL layer enabled data scientists and operations personnel to spin up complex Flink pipelines in a matter of a few hours with just basic SQL knowledge. Anyone within Uber can use PrestoSQL to query data across Pinot and other data systems (eg: Hive) in a seamless manner. Backfilling data across regions is as simple as clicking a button for executing the same query or code in a historical fashion. Moreover, these abstractions provide an extensible framework for us to evolve the underlying technologies and implement future optimizations such as tiered storage. 

\section{Future work} \label{sec:future}
Our systems continue to evolve to serve our users better. We have identified a few areas that we will invest strategically in and provide better solutions.

{\bfseries Streaming and batch processing unification} There are several use cases that demand both batch and stream processing, such as the lambda architecture and offline/real-time feature computing for machine learning. It’s common for the users to express the same processing logic twice in different languages and run on different compute frameworks. A unified processing solution will ease the development and pipeline management. 

{\bfseries Multi-region and multi-zone deployments} We are working on a multi-region-multi-zone strategy to push the scalability and reliability of our real-time data infrastructure to the next level to tolerate zone-level and region-level disasters. The biggest challenge here is to optimize data placement in order to balance data redundancy for reliability and storage cost due to excessive copies.

{\bfseries On-prem and cloud agnostic} In recent years, the ability to run system infrastructure in the cloud environment has gained a lot of importance. At Uber, we are also looking at cloud adoption and are investigating ways of converting the systems to be agnostic of data centers or cloud, so that we can move freely from on-prem to cloud.

{\bfseries Tiered storage} Storage tiering improves both cost efficiency by storing colder data in a cheaper storage medium as well as elasticity by separating data storage and serving layers. We are actively investigating tiered storage solutions for both Kafka and Pinot and collaborating closely with the open source community in this regard.

\section{Acknowledgement}
Realtime data infrastructure at Uber is an evolving architecture of multi-year effort from several teams. Many engineers, PMs and management leaders contributed to our systems, and we would like to thank them for the contributions. 

\bibliographystyle{ACM-Reference-Format}
\bibliography{references}

\end{document}